\documentclass[showpacs, aps, prb, unsortedaddress,showkeys,longbibliography]{revtex4-2}

\usepackage{amssymb}
\usepackage{latexsym}
\usepackage{amsmath}
\usepackage{graphicx}
\usepackage{subfigure}
\usepackage{braket}
\usepackage{float}
\usepackage{color,soul}
\usepackage{amsthm}
\usepackage[dvipsnames]{xcolor}

\newtheoremstyle{named}{}{}{\itshape}{}{\bfseries}{.}{.5em}{\thmnote{#3's }#1}
\theoremstyle{named}

\begin{document}
\newpage
\title{Sustainable pulling motion of an active scatterer }

\date{\today}
\author{Hossein Khodavirdi}
\author{Majid Rajabi}
\thanks{Corresponding Author}

\affiliation{Wave Research Laboratory, Department of Mechanical, Materials, and Aerospace Engineering,
Illinois Institute of Technology, Chicago, IL, 60616
USA}
\affiliation{Sustainable Manufacturing Systems Research Laboratory, Department of Mechanical Engineering,
Iran University of Science and Technology, Tehran, Tehran, Iran
}
\email{majid\_rajabi@iust.ac.ir}

\begin{abstract}
In this paper, we concern the generation of attractive net motion with respect to the location of external wave source for active spherical carriers. Here we recall that the exerted acoustic radiation force in an acoustic field, resulting from an incident wave, a scattered wave, and the radiated wave from the active carrier, can be positive, negative or zero based on the location of the object in the field. Thus in a general case, a sustainable pulling motion is not guaranteed. In this work, by considering the point that the average of the radiation force for an object, active in a single mode, over a complete wavelength is equal to the radiation force applied to a passive object (which is always positive), we put forward a technique to generate the acoustic radiation force with negative average over a complete wavelength to ensure that the net motion is attractive. The idea here is a simultaneous excitation of the object in two modes of breathing and first, with a difference in the phase of excitations. We will also show that by controlling the phase difference, the average of the force exerted on the object can be positive, negative or zero. Moreover, we show that for specific phase differences not only the average of the force is negative but also the force itself never experiences a positive value in a whole wavelength at those phase differences, which this can be a desired state to achieve a perfect negative net motion. The formulation of the paper is developed based on the velocity distribution over the surface of the sphere but an implementation method with piezoelectric actuators is suggested. Here, by setting the zero- radiation force state (i.e., radiation force function cancellation) as a reasonable criterion for determination of the required amplitude for surface velocity or the operation voltage, we calculate the averaged force vs the frequency and the phase difference. The current work, with the hope of complete maneuverability extends the concept of manipulation of smart carriers and reinforces the literature on single beam acoustic handling techniques. 

\end{abstract}


\maketitle

\section{Introduction}\label{sec:introduction}
In this paper, we introduce a simple configuration with the aim of generating attractive net motion for small carriers. The contact-free methods for pushing \cite{ARFpushing,pushing2}, pulling \cite{pullingforce1}, levitating \cite{levitation1,levitation2,levitation3}, sorting \cite{Sorting1,sorting2}, focusing \cite{trapping1,trapping2,trapping4} or in general steering of small carriers \cite{manipulation1,manipulation2,rajabikhodavirdi1} has been of interest in medical areas for applications like precise drug delivery, single cell manipulation \cite{Guo1522}, etc. More specifically, the manipulation of small objects via acoustic radiation force and torque (as two necessary drivers), resulted from the interaction of incident, scattered and radiated wave fields, on objects with different shapes has inspired a large body of scientific research. Considering the pressure gradient and force as the most fundamental concepts to develop acoustic contact-less manipulation methods, King \cite{king1934acoustic} in 1934 proposed a formulation for the acoustic radiation pressure on rigid spheres in non-viscous fluids which can be considered as a cornerstone in this literature. Hasegawa and Yosioka also propounded their formulation for the acoustic radiation pressure on an elastic solid sphere in a progressive plane wave field \cite{hasegawa1969,hasegawa1977}. Acoustic radiation force also was of concern in works like \cite{westervelt1951theory,ARFpushing,ilinskii2018acoustic}, which they found the acoustic radiation force with slightly different approaches for objects of different shapes. Moreover, major contributions were done on acoustic radiation force and torque in progressive  \cite{silva2011force,silva2014torque,silva2018progandstand} and standing wave fields \cite{kozuka2008standing,silva2018progandstand} for different object shapes \cite{hasegawa1988cylind,lima2020nonlinear,jerome2021acoustic}.\par


Negative acoustic radiation force has been discussed as a necessary degree of freedom to achieve the complete acoustic maneuverability of objects in works like \cite{Marston2006,Zhangmarston2011,ZhangMarston2012,xu2012transversally,marston2007negative,gong2021non,gong2020far}.  Considering the incidence of a simple progressive plane wave field with a passive object, an intuitive deduction will be the pushing effect on the passive object which can be justified by referring to the transfer of linear momentum from the wave to the object. Also, because of the wave-object interaction there will be a scattered field, which because of that, a recoil force will be applied to the object. In this case, for a forward dominant scattered field the induced force will be in the opposite direction \cite{acousticpullingplanewave,opticalpullingforce}. Thus, the endeavors on generating negative (pulling) motion are either on maximizing the forward scattering to have a big intensity of pulling force or minimizing the projection of incident wave momentum in the forward direction to have a small pushing force. The latter has been addressed in works by  Marston and Zhang \cite{Marston2016,ZhangMarston2016} in which they showed that for a passive object placed in a progressive plane wave field, it is not possible to have negative absorbed energy or negative acoustic radiation force. Nevertheless, due to the smaller forward projection of incident momentum for Bessel beams comparing to the simple plane wave case, the same passive object in a focused acoustic field like Bessel beam experiences negative acoustic radiation force at specific frequencies and cone angles \cite{Marston2006}. \par 

For the generation of pulling force, using an energy-wise open scatterer is the alternative to using complex wave fields \cite{Pullingongainmedium, PRLNegForce,KHODAVIRDI2022104399}. Here, the goal of using active objects is to obtain specific states on which the interference between radiation from the active object and the forward scattering due to the incident wave be constructive. In other words, by reinforcing the waves in the forward direction (forward scattered wave+radiation), a bigger backward momentum will be applied to the object itself. In this paper, after introducing the new technique, we will show that for specific states of activeness, the forward direction waves can be so big that guarantees not only a pulling force but also a sustainable pulling motion. The recent efforts on developing the concept and idea of active carriers on the path of promoting the acoustic handling technique, led to the introduction of the new concepts of acoustically activated bodies as self-motile carriers \cite{mojahed2018self}, self-activated carriers manipulated with the monochromatic incident beam \cite{Mojahedrajabi1,Mojahedrajabi2,Mojahedrajabi3,MojahedRajabi4,MojahedRajabi5} , or externally activated bodies with compound fields \cite{yu2018negativelaser,yu2019laser} and just recently, it is shown that how 3D steering may be achieved via a bi-polar configuration of the internal excitation of a spherical active carrier \cite{rajabikhodavirdi1}. Although the empirical aspect of the design and fabrication of active carriers using piezo ceramics has not been seen by the authors yet, there are many published papers based on practical experiments on acoustic manipulations of passive particles using different acoustic fields \cite{Drinkwater20141,levitation2,levitation3}. Moreover, Hai-Qun Yu proposed a method \cite{yu2019laser} which seems feasible to be tested practically which is about exciting liquid spheres by a laser beam instead of using piezo ceramics to generate negative acoustic radiation force, albeit that work is again a theoretical study.\par
In the present work, another idea of manipulation strategy is pursued focusing on the concept of generating net negative radiation force independent of the location of the carrier. The active carrier is stimulated on its two first modes of vibration (i.e., mono-pole and dipole modes) with specified phase difference and the same frequency. The following sections start with a problem definition in section (\ref{sec:probdefinition}) and the formulation section comes after it starting with the acoustic field equations in sub-section A along with the calculations of scattered coefficients in sub-section B. Section (\ref{sec:formulation}) continuous with two  sub-sections, first a mathematical manipulation on the acoustic radiation force formulation in sub-section C and then a quick review on the formulae for a piezo-elastic structure in sub-section D. Section (\ref{sec:formulation}) ends with a summary about the dynamics of a moving particle in a fluid in sub-section E. In the results and discussion section, by giving a numerical example, first it is shown that the formulations are valid for the case of a single-mode activated sphere, then Fig. (\ref{Yzonemodeactivevsphi0ka4.1}) shows that for objects which are activated in a single mode, the average of the force over the location of the object is always positive and equal to the acoustic radiation force (ARF) acting on a passive object. Also in section (\ref{sec:RD}), it is shown that for a wide range of phase differences between two excitation modes of the carrier, at some specified frequencies, the net negative radiation force is achievable.  In the continuation of the result section, by considering the dynamics of the carrier in the host fluid medium and assuming the linear time variation of the phase of incident wave field, it is shown that four states of the radiation force are possible:
\begin{itemize}
    \item The positive average of radiation force (RF) with always positive values of RF. 
    \item The positive average of radiation force with positive/zero/negative values of RF on a wavelength and induction of stable/unstable rest states.
    \item The negative average of radiation force with always negative values of RF.
    \item The negative average of radiation force with positive/zero/negative values of RF on a wavelength and induction of stable/unstable rest states.
\end{itemize}

Taking into account the dynamics of the active carrier in the host medium and by keeping in mind the existence of the drag force, the history force and the added mass effects \cite{happel2012low}, the time dependency of traveling velocity of the carrier is calculated and it is observed that for the rate of phase change greater than a (frequency dependent) minimum value, the traveling velocity of the carrier approaches the stable value, backward the incident wave field direction.

\section{Problem Definition}\label{sec:probdefinition}

\begin{figure}[!htb]
\includegraphics[scale=0.3]{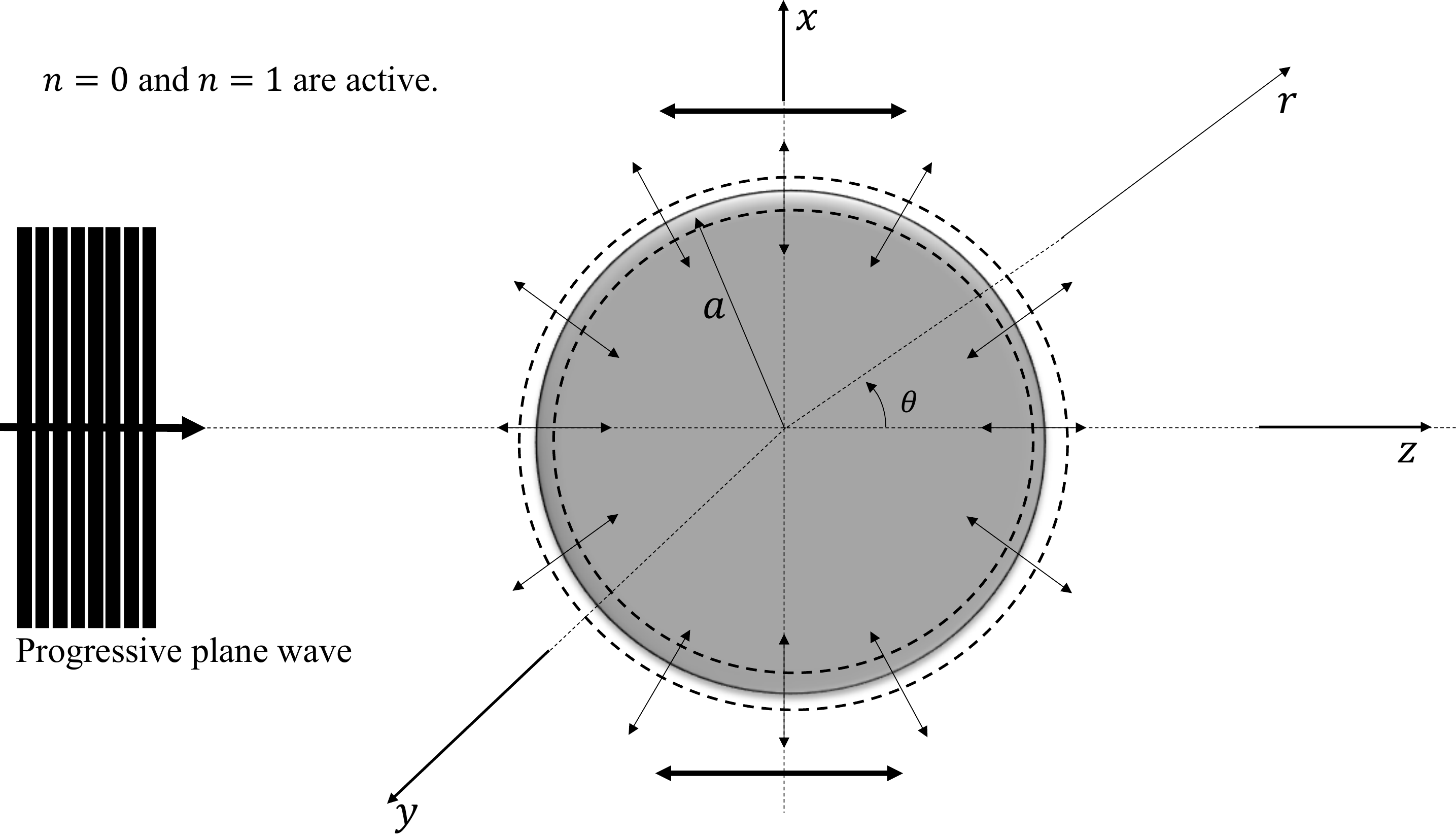}
\centering
\caption{ Configuration of The problem. It represents the interaction of a progressive plane wave and a sphere active in its zero and first modes.}
\centering
\label{Config}
\end{figure}

In this section we give a clear definition of the problem configuration. Fig. (\ref{Config}) is showing a carrier active in its zero mode with the surface velocity of $v_0$ and the first mode, with $v_1$ as its surface velocity. Here, the amplitudes of $v_0$ and $v_1$ are equal but a varying phase between the two modes always exist as, $\Delta\phi$, which let us write, $v_0=v_{amp}e^{-i\phi_0}$ and $v_1=v_{amp}e^{-i\left(\phi_0+\Delta\phi\right)}$. In this paper, once we develop a formulation without considering the equations for a piezo-elastic structure sphere which addresses the mechanism of activating the body and simply find the scattering coefficients by directly assuming a desired velocity profile on the surface of the object (like $v_0$ and $v_1$), and once review the implementation of the ideas for a piezo-elastic sphere and present results for both perspectives.

As is shown in Fig (\ref{Config}), to represent vibrations in the breathing mode ($n=0$), a homogeneous cyclic radial velocity excitation is required. Moreover, for the first mode vibration, cosine distribution of the radial velocity guarantees oscillations in the z-direction and the total radial velocity of the surface can be written as $\boldsymbol{v}=(v_0+v_1\cos\theta)e^{-i\omega t}\hat{e}_r$. Fig. (\ref{Config}) also represents an incident plane wave propagating towards the elastic sphere with the outer radius of $a$ which is active in zero and first modes with a phase deference of $\Delta \phi$ between the modes. 



\section{Formulations}\label{sec:formulation}
In the following sub-sections, we first recall the acoustic field equations for an inviscid compressible fluid in an adiabatic condition. Then, by directly assuming velocity profiles for the surface of the object as a result of an arbitrary method of actuation of the sphere, using the continuity condition at the interface of the fluid and the scatterer, the scattered coefficients will be calculated and they will be represented in a way to explicitly show the phase of object's vibrations. In addition, the acoustic radiation force function along with a classification on its terms are represented, to highlight the importance of phase difference between the two modes of sphere vibrations. The formulation section also has a summary about the equations of a piezo-elastic structure, to introduce one of way of implementing an active carrier and and its last part goes for the dynamics of a moving object in the fluid to see how the object follows the changes in the phase of incident wave field. In the following parts, bold symbols are showing vectors or matrices and $\braket{}$ indicates either an inner product with a weight function of $1$ or an averaging operation, depending on the context. 

\subsection{Acoustic Field}
Following the classical approach, the generalized form of Newton's second law equation, here called Euler's equation, for a compressible fluid, flowing at small velocity can be written like \cite{foundationofacoustics}:
\begin{equation}
    \label{Euler}
    -\boldsymbol{\nabla}p=\rho_0\frac{\partial\boldsymbol{v}}{\partial t}
\end{equation}
where, $p$ is the pressure, $\rho_0$ is the mean value of the fluid density, and $\boldsymbol{v}$ is the velocity vector of the fluid. The continuity equation for a non-viscous fluid in an adiabatic condition can be shown as:

\begin{equation}
    \label{continuity}
    -\boldsymbol{\nabla}.\boldsymbol{v}=\frac{1}{\rho_0c^2}\frac{\partial p}{\partial t}
\end{equation}
Now, by combining Eqs. (\ref{Euler}) and (\ref{continuity}) and using the technique of Helmholtz decomposition in a non-viscous fluid, the time-independent form of wave equation can be written like:
\begin{equation}
    \label{wave}
    \left(\boldsymbol{\nabla}^2+k^2\right)\varphi(\boldsymbol{r})=0
\end{equation}
where $k$ is the wave number which here can be represented as $k=\frac{\omega}{c}$ and $\varphi(\boldsymbol{r})$ is the velocity potential field. 
Considering the origin at the center of the sphere, the solution to the wave equation for the incident field for a case with a symmetry in azimuthal angle can be written like:

\begin{equation}
    \label{incwave}
    \varphi_{inc}(r,\theta,\omega,t)=\varphi_0\sum_{n=0}^{\infty}\left(2n+1\right)i^n j_n(kr)P_n(\cos \theta) e^{-i\omega t}
\end{equation}
Also, considering the Sommerfeld radiation condition \cite{junger1986sound}, the solution to the wave equation for the scattered field will be:
\begin{equation}
    \label{sccwave}
    \varphi_{sc}(r,\theta,\omega,t)=\varphi_0\sum_{n=0}^{\infty}A_n\left(2n+1\right)i^n h_n^{(1)}(kr)P_n(\cos \theta) e^{-i\omega t}
\end{equation}
In Eqs. (\ref{incwave} and \ref{sccwave}), $A_n$ are modal scattering coefficients, $P_n(\cos \theta)$ are the Legendre polynomials, $j_n(kr)$ is the spherical Bessel function and $h_n^{(1)}(kr)$ is the spherical Hankel function \cite{mathhandbook}. The total radial velocity in the fluid results from the incident and scattered wave also will be:
\begin{equation}
    \label{totalvel}
    v_{r}^{f}(r,\theta,\omega,t)=\varphi_0k\sum_{n=0}^{\infty}\left(2n+1\right)i^n \left(A_n h_n^{(1)'}(kr)+ j_n(kr)^{'}\right)P_n(\cos \theta) e^{-i\omega t}
\end{equation}
where, $(.)'$ means the first derivative with respect to $r$ and the superscript of $f$ means the velocity in the fluid medium.
\subsection{Modal Scattering Coefficients}\label{modal}
For a given set of $r$, $\theta$, $\omega$ and $t$ the only unknown parameter in the total velocity potential and velocity itself, will be $A_n(\omega)$ which are scattering coefficients and they carry the properties of the scatterer with themselves. Here, because of the simple physics of the system, a continuity consideration at $r=a$ for the radial velocity is enough to find all the scattering coefficients. Considering the point that the wave solution is represented as an expansion over a complete orthogonal basis of Legendre polynomials, we expand the velocity of the surface of the active carrier on the same basis, like:
\begin{equation}
    \label{velocityexpansion}
    v_{r}^{obj}(\theta,\omega,t)=\sum_{n=0}^{n=\infty}v_n P_n(\cos\theta)e^{-i\omega t}
\end{equation}
where, $v_n$ are the modal coefficients of velocity in the expansion. At this point,  by recalling from section (\ref{sec:probdefinition}) that the total velocity on the surface of the object due to the prescribed excitation (applied by either a Piezoelectric ceramic, a laser irradiation, a chemical process ,etc) can be written like, $\boldsymbol{v}=(v_0+v_1\cos\theta)e^{-i\omega t}\hat{e}_r$, and also by noting the orthogonality condition between Legendre polynomials in the form of $\braket{P_n,P_m}=\frac{2}{2n+1}\delta_{nm} $, we can write the modal coefficients of the velocity like: 

\begin{equation}
\label{veloc}
  v_{n} =
    \begin{cases}
      v_{amp}e^{-i\phi_0} & n=0\\
      v_{amp}e^{-i\left(\phi_0+\Delta\phi\right)} & n=1\\
      0 & \text{otherwise}
    \end{cases}
\end{equation}
Having the modal coefficients of the velocity in hand, one can use Eq. (\ref{totalvel}) and Eq. (\ref{velocityexpansion}) and conduct the continuity of the radial velocity at the interface of the fluid and sphere and find the scattering coefficients as:
\begin{equation}
\label{scatteringcoef}
  A_{n}(\omega) =
    \begin{cases}
      \frac{v_n+\varphi_0 k (2n+1)i^n j_{n}^{'}(ka)}{-\varphi_{0}k(2n+1)i^nh_{n}^{(1)'}(ka)} & n=0 \& n=1\\
      -\frac{j_{n}^{'}(ka)}{h_{n}^{(1)'}(ka)} & \text{otherwise}
      
    \end{cases}
\end{equation}
Eq. (\ref{scatteringcoef}) can also be expressed like:

\begin{equation}
\label{scatteringcoef2}
  A_{n}(\omega) =
    \begin{cases}
      Z_{1}^{n}+v_{n}Z_{2}^n & n=0 \& n=1\\
      Z_1^n & \text{otherwise}
      
    \end{cases}
\end{equation}
where, $Z_1^n=-\frac{j_{n}^{'}(ka)}{h_{n}^{(1)'}(ka)}$ and $Z_2^n=-\frac{1}{\varphi_{0}k(2n+1)i^nh_{n}^{(1)'}(ka)}$. Eq. (\ref{scatteringcoef}) indicates that the scattering coefficients for a passive object (like the case for $n>1$) are independent of the amplitude and phase of the incident wave field which this means that the exerted acoustic radiation force is also independent of the location of the object with respect to the incident wave source. Nevertheless, the existence of both complex amplitude of the velocity potential for the incident wave ($\varphi_0$) and the modal coefficients of the velocity of the object due to its activeness ($v_n$), in the same term implies that the scattering coefficients at modes equal to the modes of activeness of the sphere ($n=0 \& 1$) will be a function of both amplitude and phase of incident wave field and the velocity of excitation. In other words, the phase of scattering coefficients $A_0$ and $A_1$ are roughly functions of the difference in the phase of the prescribed velocity of sphere's surface and the phase of incident wave means, $\angle A_0=\text{func}\left(\angle v_0-\angle \varphi_{inc}\right)$ and $\angle A_1=\text{func}\left(\angle v_1-\angle \varphi_{inc}\right)$. Here, to simplify the notation and to explicitly show the effect of object's location in formulations, one can set a constant prescribed phase for either the modal velocities or the incident wave and change the other phase in order to see the effect of objects location on the exerted force. In this paper, we consider the phase of incident wave as a constant number and so any change in the phase of modal velocity of $n=0$ can be interpreted as the location of the object with respect to the wave source.   

\subsection{Acoustic Radiation Force}\label{AcousticRF}
In a non-linear regime, the force exerted on an object illuminated by a wave field is non-zero. For the case of progressive plane wave, we review the formulation for the  so-called Acoustic Radiation Force exerted on a sphere with the outer surface at equilibrium. The time-averaged radiation force function in the z-direction can be represented as \cite{Mojahedrajabi1}:

\begin{equation}
\label{ARF1}
\braket{F}_t=E_{inc}S_c Y_z
\end{equation}

where $E_{inc}=\rho k^2 \varphi_0^2/2$  is the incident wave energy density and $S_c$  is cross-sectional area of the spherical object.  Moreover, $Y_z$ is the non-dimensional form of the acoustic radiation force in z-direction as \cite{Force1,forcexyz}:
\begin{equation}
    \label{ARF2}
    Y_z=\frac{-4}{\left(ka\right)^2}\sum_{n=0}^{\infty}\left(n+1\right)\left[\alpha_n+\alpha_{n+1}+2\left(\alpha_n\alpha_{n+1}+\beta_n\beta_{n+1}\right)\right]
\end{equation}
where  $\alpha_i$ and $\beta_i$  are real and imaginary parts of the scattering coefficients, respectively. The Eq. (\ref{ARF2}) in conjunction with the scattering coefficients obtained in Eq. (\ref{scatteringcoef2}) give an insight about the effect of vibrations of the sphere on the resulted acoustic radiation force. In other words, by keeping in mind the different expressions for $A_n$ at $n=0\&1$ and $n>1$, one can classify the expressions in Eq. (\ref{ARF2}) as two main terms of $Y_z^{active}$ which can be expressed as $Y_z^{active}=\text{func}(v_{amp},\phi_0,\Delta\phi,k)$ and $Y_z^{passive}$ which is only a function of frequency. In previous sub-section, based on the assumption made in this paper, $\phi_0$ can be interpreted as the location of the particle with respect to the wave source. After a minor manipulation in Eq. (\ref{scatteringcoef2}) and by using Eq. (\ref{veloc}), for $n=0$ we have :

\begin{equation}
\begin{aligned}
    \label{alphabeta0}
    \alpha_0=\Re Z_1^0+\Re Z_2^0v_{amp}\cos\phi_0+\Im Z_2^0v_{amp}\sin\phi_0\\
        \beta_0=\Im Z_1^0+\Im Z_2^0v_{amp}\cos\phi_0-\Re Z_2^0v_{amp}\sin\phi_0
\end{aligned}
\end{equation}
and for $n=1$:
\begin{equation}
\begin{aligned}
    \label{alphabeta1}
    \alpha_1=\Re Z_1^1+\cos\phi_0v_{amp}\left(\Re Z_2^1\cos\Delta\phi+\Im Z_2^1\sin\Delta\phi\right)+\sin\phi_0v_{amp}\left(\Im Z_2^1\cos\Delta\phi-\Re Z_2^1\sin\Delta\phi\right)\\
        \beta_1=\Im Z_1^1+\cos\phi_0v_{amp}\left(\Im Z_2^1\cos\Delta\phi-\Re Z_2^1\sin\Delta\phi\right)-\sin\phi_0v_{amp}\left(\Im Z_2^1\sin\Delta\phi+\Re Z_2^1\cos\Delta\phi\right)
\end{aligned}
\end{equation}
Where, $\Re$ ($\Im$) before any parameter means the real (imaginary) part of that parameter. Now, by substituting Eq. (\ref{alphabeta0}) and (\ref{alphabeta1}) into Eq. (\ref{ARF2}) and after some manipulations, the expression for the non-dimensional force function can be re-written in a way which is more consistent with our goal, like:
\begin{equation}
    \label{ARF3}
    Y_z=\frac{-4}{\left(ka\right)^2}\left(A\cos\phi_0+B\sin\phi_0+C\sin\phi_0\cos\phi_0+D\sin^2\phi_0+E\cos^2\phi_0\right)+Y_z^{passive}
\end{equation}
where, 
\begin{multline}
    \label{Ypassive}
    Y_z^{passive}=\frac{-4}{\left(ka\right)^2}\bigg\{\alpha_2+\sum_{n=2}^{\infty}\left(n+1\right)\left[\alpha_n+\alpha_{n+1}+2\left(\alpha_n\alpha_{n+1}+\beta_n\beta_{n+1}\right)\right]+\\
    \Re Z_1^0+2\Re Z_1^0\Re Z_1^1+2\Im Z_1^0\Im Z_1^1+2\alpha_2\Re Z_1^1+2\beta_2\Im Z_1^1\bigg\}
\end{multline}
and the coefficients $A$, $B$, $C$, $D$ and $E$ are all functions of frequency and $\Delta\phi$ and they are presented in the appendix (section \ref{appendix1}). A closer look into Eqs. (\ref{ARF3}) and (\ref{ARF2}) along with keeping in mind the expressions for $\alpha_0$, $\beta_0$, $\alpha_1$ and $\beta_1$ leads to a level of deeper insight into the idea of activating the object in two modes with a shift in their phases. In other words, one can see that in the case of single mode excitation of the sphere, the acoustic radiation force function will be only a function of $\cos\phi_0$ and $\sin\phi_0$ which by taking the average of the force over a complete wave length, the only non-zero term will be $Y_z^{passive}$ which is always positive \cite{UltrasonicMojahed}. Thus, a double mode excitation of the sphere with a varying phase difference brings about a level of flexibility to manipulate coefficients $D$ and $E$ in Eq. (\ref{ARF3}) to generate negative average for the acoustic radiation force over a complete wavelength. In this way, one can make sure that the active object is always experiencing a pulling motion everywhere in the wave field. 
\subsection{Implementation with Piezo-elastic Structure}\label{piezoimplementation}
\begin{figure}[!htb]
\includegraphics[scale=0.3]{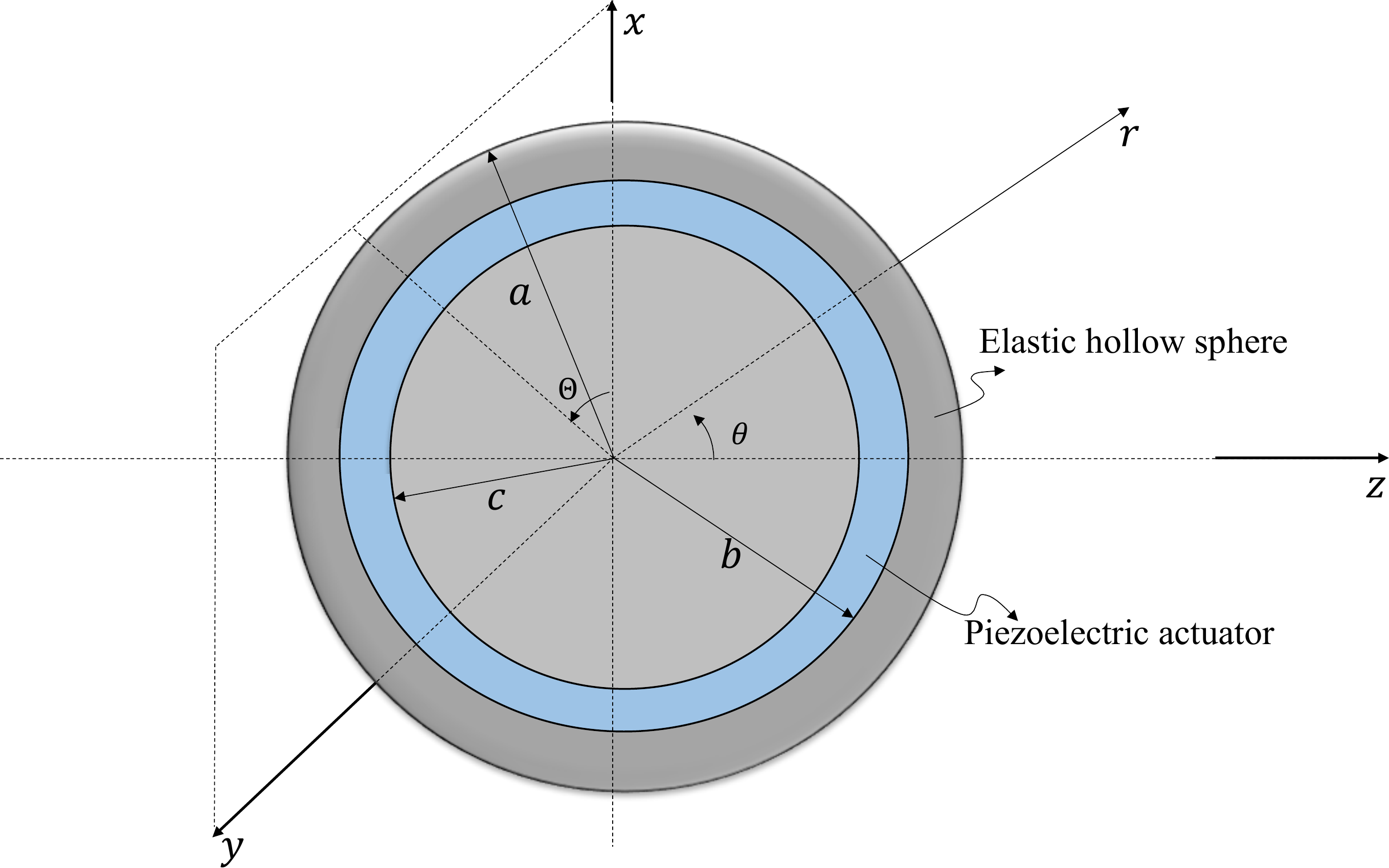}
\centering
\caption{An elastic hollow sphere equipped internally with a layer of piezoelectric actuator.}
\centering
\label{piezosteel}
\end{figure}

Following the developments in previous subsections, one might ask whether there is any more realistic way to model the suggested method in this paper. In other words, how would be the formulation if we start from a more preliminary step and think about a mechanism to excite the sphere in its two first modes with a difference in the phase. Here, we take advantage of the method introduced by Scandrett \cite{Scandrett} for scattering of the wave from a submerged bi-laminate spherical shell which is made of an elastic hollow sphere which is internally coated with a layer of Piezo-ceramic actuator (Fig. (\ref{piezosteel})). Here, we insist that the goal of this section is not to present a detailed formulation for the piezo-elastic structure, and we only give a short summary of the steps and for further readings, we refer the reader to the literature \cite{Scandrett,Mojahedrajabi1} and the appendix (\ref{elastodyn}).    

Starting with the constitutive equations in frequency domain for the elastic and Piezo-electric, we have \cite{Mojahedrajabi1,chen2002free,chen20023d,chen2000vibration}:
\begin{eqnarray}
\label{Constitutive} 
\boldsymbol{\sigma}=\boldsymbol{c}\boldsymbol{s};\quad \boldsymbol{\Sigma}=\boldsymbol{C}\boldsymbol{S}-\boldsymbol{\chi}\boldsymbol{\Phi}; \quad \boldsymbol{\Delta}=\boldsymbol{eS}+\boldsymbol{\Lambda}\boldsymbol{\Phi}
\end{eqnarray}
in which, $\boldsymbol{c}$ and $\boldsymbol{C}$ are stiffness tensors for the elastic layer and the piezo respectively and $\boldsymbol{e}$ is the piezoelectricity matrix. Moreover,   $\boldsymbol{\sigma}$ and $\boldsymbol{s}$ are the stress and strain vectors for the isotropic elastic casing, $\boldsymbol{\Sigma}$, $\boldsymbol{S}$ and $\Delta$ are stress, strain and electric field for the piezoelectric material and $\boldsymbol{\chi}$ and $\boldsymbol{\Lambda}$ are two matrix operators. The expressions for the symbols in Eq. (\ref{Constitutive}) are represented in section (\ref{appendix2}). In the absence of an external force and for the case of no free charge density, two sets of equations of motion (dynamics of the structures) along with the Gaussian equation of electric equilibrium can be represented as:
\begin{equation}
    \label{EOM}
    \begin{aligned}
        \mathbf{Q}\boldsymbol{\sigma}=\rho_{e}\boldsymbol{\Ddot{u}}\\
        \mathbf{Q}\boldsymbol{\Sigma}=\rho_{p}\boldsymbol{\Ddot{U}}\\
        \nabla_2\Delta_r+\Delta_r+\left(\frac{1}{\sin \theta}\right)\frac{\partial}{\partial\theta}\left(\Delta_{\theta}\sin \theta\right)+\left(\frac{1}{r\sin \theta}\right)\frac{\partial}{\partial\Theta}\left(\Delta_{\Theta}\right)=0
    \end{aligned}
\end{equation}
where, $\boldsymbol{\Ddot{u}}$ and $\boldsymbol{\Ddot{U}}$ are second time derivatives of displacement vectors in elastic layer and piezoelectric respectively and $\rho_e$ and $\rho_p$ are the corresponding densities. Also, as $\phi$, $\Phi$ and $\varphi$ are all used for other situations through the paper, here we used big theta ($\Theta$) to represent azimuthal angle in a spherical coordinates. From this point, according to the method used in \cite{chen20023d,Mojahedrajabi1}, by keeping in mind the constitutive relations in Eq. (\ref{Constitutive}) along with the equations in Eq. (\ref{EOM}) one can apply the change of variables introduced in \cite{chen20023d}, use a laminate model introduced in \cite{tanigawa1995some,chen20023d,Mojahedrajabi1}, apply the solid-solid/fluid-solid continuity condition in association with mechanical/electrical boundary conditions and use the expansion of state variables on Legendre polynomials and after a tedious and long manipulation find the scattering coefficients in the following form:

\begin{equation}
\label{scatteringcoefvolt}
  A_{n}(\omega) =
    \begin{cases}
      {\mathcal{Z}}_{1}^{n}+\Phi_{n}{\mathcal{Z}}_{2}^n & n=0 \& n=1\\
      {\mathcal{Z}}_1^n & \text{otherwise}
      
    \end{cases}
\end{equation}
where ${\mathcal{Z}}_{1}^{n}$ and ${\mathcal{Z}}_{2}^{n}$ are the modal transfer functions of the system present in the appendix and for a detailed version readers are referred to \cite{Mojahedrajabi1}. Also, $\Phi_n$ is the non-dimensional modal voltage amplitude applied to the piezoelectric actuator which is obtained from the modal expansion of the imposed electric potential amplitude, in the form:

\begin{equation}
    \label{volt1}
    l(\theta,\omega)=\left(\frac{e_{33}a}{\epsilon_{33}}\right)\sum_{n=0}^{\infty}\Phi_n P_n\left(\cos\theta\right)
\end{equation}

${\empty}$
which here, to prevent confusion between a prescribed velocity and a prescribed voltage, we used $l$ to represent voltage. Similar to the solution based on displacement excitation, to excite the sphere on the breathing and first mode with a phase difference, we choose $l=\left(l_0+l_1\cos\theta\right)e^{-i\omega t}$ as the distribution of the prescribed voltage. Here, $l_0$ and $l_1$ are complex amplitudes of the voltage and they are $l_0=l_{amp}e^{-i\phi_0}$ and $l_1=l_{amp}e^{-i\left(\phi_0+\Delta\phi\right)}$ respectively. Using the orthogonality of the Legendre polynomials, we can right:
\begin{equation}
\label{scatteringcoefvolt2}
  A_{n}(\omega) =
    \begin{cases}
      \boldsymbol{\mathcal{Z}}_{1}^{0}+\frac{\epsilon_{33}}{e_{33}a}l_{amp}e^{-i\phi_0}\boldsymbol{\mathcal{Z}}_{2}^0 & n=0\\
      \boldsymbol{\mathcal{Z}}_{1}^{1}+\frac{\epsilon_{33}}{e_{33}a}l_{amp}e^{-i\left(\phi_0+\Delta\phi\right)}\boldsymbol{\mathcal{Z}}_{2}^1 & n=1\\
      \boldsymbol{\mathcal{Z}}_1^n & \text{otherwise}
      
    \end{cases}
\end{equation}
From this point on, due to the vast similarity between Eq. (\ref{scatteringcoefvolt2}) and (\ref{scatteringcoef2}) all the developed formulations for the non-dimensional acoustic radiation force and its average for the case of displacement excitation are valid except that the expressions of coefficients $A,B,..$ are slightly different.  

\newpage
\subsection{Dynamics of the Scatterer}
In most of the original works done in acoustic manipulation area of research, the attentions were mostly on the steady solutions of the problems to improve and complete the concepts along with adding mathematical methods to enrich the topic of acoustic manipulation. Nevertheless, to cultivate and support the concept, the correlation between the rate of change in particle phase and the velocity of carrier due to the effect of acoustic radiation force, Stokes drag, added mass force and Basset force is investigated. The equation of motion of a particle in viscous fluid is like \cite{happel2012low},
\begin{equation}
    \label{Hydrodynamics}
    m_c\frac{dV_c}{dt}=-6\pi\eta a V_c-\frac{1}{2}m_f\frac{dV_c}{dt}-6\pi\eta a^2 \int_0^td\tau \frac{\frac{dV_c}{d\tau}}{\left[\pi\nu\left(t-\tau\right)\right]^{\frac{1}{2}}}+F_{acoustic}
\end{equation}

The first, second and third terms in right hand side of Eq. (\ref{Hydrodynamics}) are Stokes drag force, added mass force and history force respectively and the last term here is the acoustic radiation force. For this equation, $V_c$  is the velocity of carrier, $\eta$  is  the dynamic viscosity, $\nu$ is the kinematic viscosity, $m_c$ is the mass of carrier (scatterer) and $m_f$  is the mass of the displaced fluid.  Here, we used Laplace transform to solve Eq. (\ref{Hydrodynamics}) and find the expression for velocity of particle.  Due to the point that Basset force has a meaningful magnitude just when the body accelerates with high rate and considering the point that its existence in the equation makes the mathematical manipulation cumbersome in vain, along with the point that we are just looking for an approximate behavior of the body, we neglect the effect of history force \cite{johnson1handbook}. After implementing Laplace transform, we will have:
\begin{equation}
    \label{Hydrodynamics2}
    \left[\left(m_c+\frac{m_f}{2}\right)s+6\pi\eta a\right]\overline{V}_c=\overline{F}_{acoustic}
\end{equation}
Here, ($\overline{.}$) means the transformed function.  The final expression for the velocity of particle will be $V_c=f*g=\int_0^t F(t-\tau)g(\tau)d\tau$, where ($*$) is the convolution product, $g=\mathcal{L}^{-1}\left\{1/\left[\left(m_c+m_f/2\right)s+6\pi\eta a\right]\right\}$, $F$ is the acoustic radiation force and $\mathcal{L}^{-1}$ means the Laplace inverse of the function comes after it.

\section{Results and Discussions}\label{sec:RD}
\begin{figure}[htp]
\centering
\includegraphics[scale=.22]{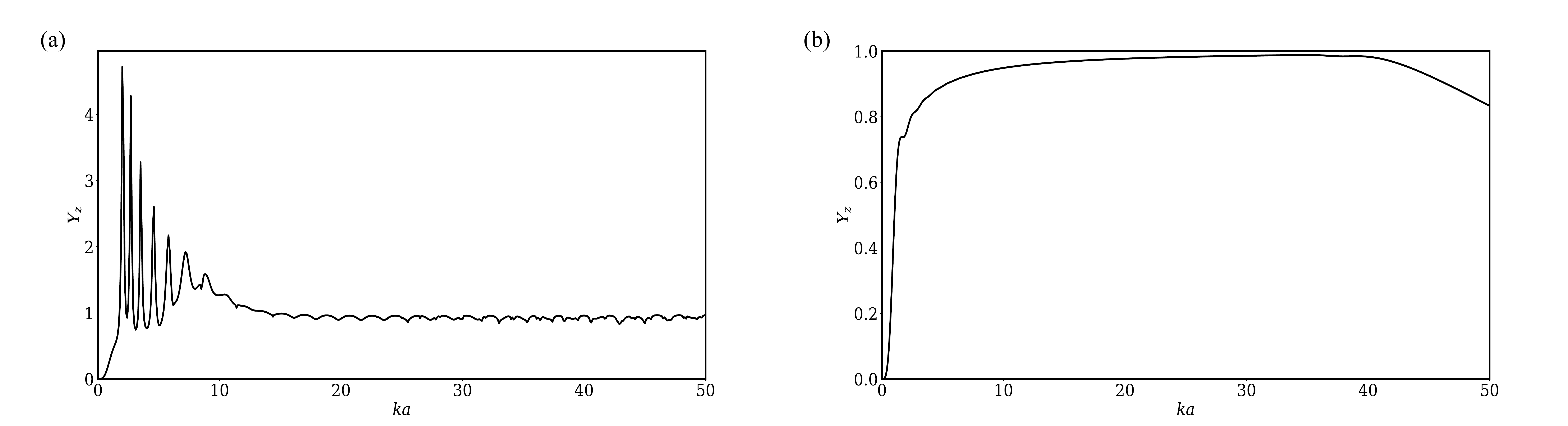}
\caption{Acoustic radiation force in z-direction vs. non-dimensional frequency for (a) a passive elastic sphere and (b) a passive rigid sphere.}
\label{passiveforcesvsfreq}
\end{figure}

To visualize the result of formulations and methods introduced in this paper, two numerical examples are provided here. Both of the examples are considering same physical problem but in one of them, the actuation of the sphere is defined in terms of the surface velocity of the sphere and the cause of the actuation is not considered (scattering coefficients according to formulations in section (\ref{modal})), while in the other example the scattering coefficients are calculated for a stainless steel spherical shell which is internally coated with a layer of $0.1$ mm PZT4 piezoelectric  using the formulation in section (\ref{piezoimplementation}). For the both examples, the surrounding fluid is assumed to be water with the pressure amplitude of ($p_{inc}$) equal to $100$ kPa along with the sound speed of $1497$ m/s. On the assumed condition, the fluid density is $997$ kg/$m^3$. Here, the outer radius of the object is equal to $1$ mm and the other radii in Fig. (\ref{piezosteel}) are as, $b=0.9$ mm and $c=0.8$ mm. The material properties for the stainless steel and PZT4 are mentioned in the appendix (\ref{MP}). For assurance, first we examine our codes, in both examples, and check whether the figures of non-dimensional acoustic radiation force and scattering coefficients vs. $ka$ (non-dimensional frequency) for a passive scatterer match with the literature or not. In the following figures, the operational frequency is chosen to be in the range of $20$kHz$<f<10$MHz which is within the frequency range of practical ultrasonic transducers and for $a=1$mm it maps to $0.1<ka<50$. Fig. (\ref{passiveforcesvsfreq}) shows the non-dimensional acoustic radiation force in z-direction vs. $ka$ for, (a) an elastic passive sphere and (b) for a passive rigid sphere, both illuminated by a progressive plane wave. In the both cases the figures are matched with the literature \cite{Mojahedrajabi1,UltrasonicMojahed,silva2011force}. To avoid unnecessary figures, the plots for scattering coefficients vs. $ka$ are not mentioned here.

To implement the methods, for both cases of activating the body using a piezo-elastic structure or just by referring to its surface velocity, we need to determine appropriate amplitudes for the voltage and surface displacement. Following the method in \cite{rajabikhodavirdi1,Mojahedrajabi1}, the calculation of minimum cancellation voltage (or surface displacement) can be a reasonable place to start with. In fact, to calculate the minimum cancellation voltage for a single mode active body one can do some manipulations on Eq. (\ref{ARF2}) and write it in the form of $Y_z=A\overline{\alpha}+B\overline{\beta}+C$, where $\overline{\alpha}$ and $\overline{\beta}$ are the real and imaginary parts of the voltage. Now, by setting the equation equal to zero, the minimum necessary voltage amplitude for a no-force condition will be the radius of the origin-centred circle tangent to the line $A\overline{\alpha}+B\overline{\beta}+C=0$ \cite{Mojahedrajabi1}. The amplitude and phase of minimum surface displacement can be calculated through a similar process. 
\newpage

\begin{figure}[htp]
\centering
\includegraphics[scale=.22]{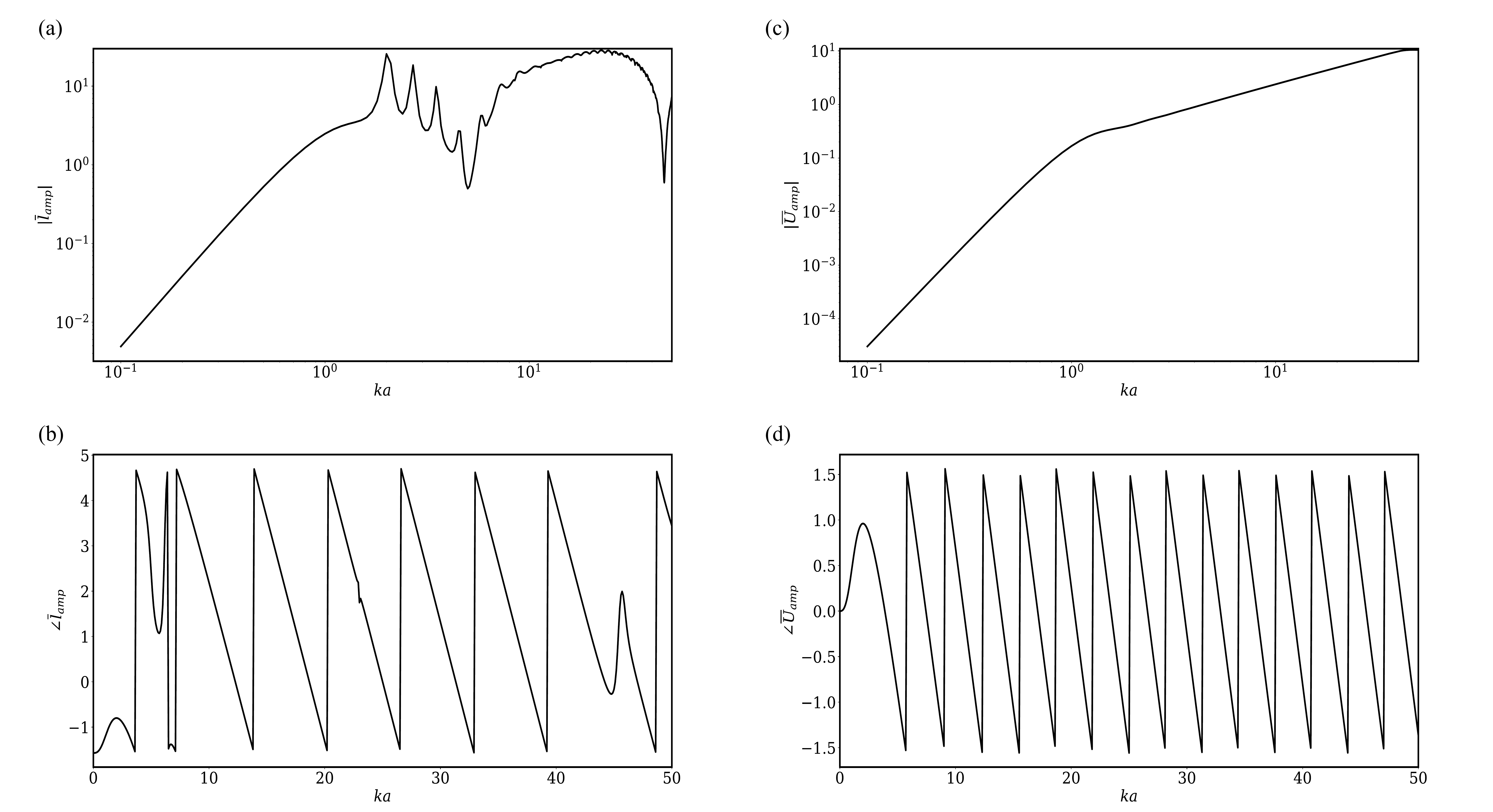}
\caption{(a) and (b) are the minimum normalized cancellation voltage and its phase vs. non-dimensional frequency respectively, for an elastic sphere. (c) and (d) are the minimum normalized surface displacement and its phase vs. non-dimensional frequency respectively, for a rigid sphere. (a) and (c) are represented in logarithmic scale.}
\label{minvoltanddisp}
\end{figure}
On the left side of Fig. (\ref{minvoltanddisp}), plots (a) and (b) are showing the minimum normalized cancellation voltage amplitude ($\overline{l}_{amp}=e_{33}l_{amp}/(a|p_{inc}|)$) and phase respectively vs. non-dimensional frequency. Plots (c) and (d) on the right are also showing the minimum required normalized surface displacement amplitude ($\overline{U}_{amp}=v_{amp}/(\omega|U_{inc}|)$) and phase respectively to cancel the acoustic radiation force in z-direction, vs. $ka$. The results of amplitudes (normalized voltage or displacement) from Fig. (\ref{minvoltanddisp}) can be multiplied by a real number, like $\gamma>1$, to become the operational amplitude for the rest of the results. Here we chose $\gamma=3$ to make sure that the object can potentially experience all negative, zero and positive force states at each frequency if it is activated in single mode. The chosen amplitudes in this part, will be used as the operational amplitudes in activation of the body in both breathing and first modes. 

\begin{figure}[htp]
\centering
\includegraphics[scale=.22]{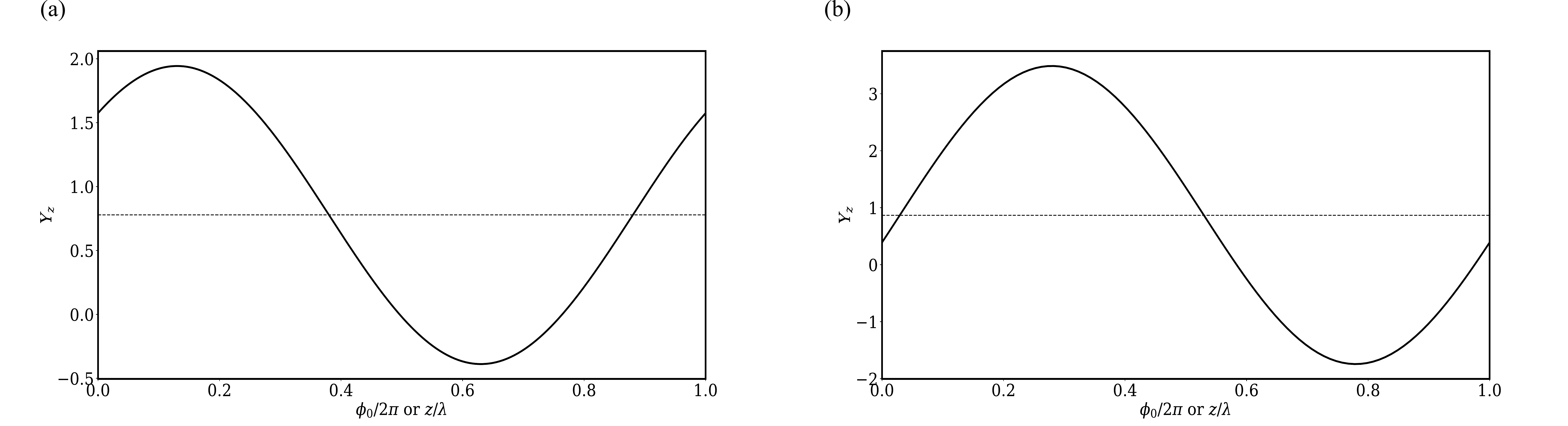}
\caption{At $ka=4.1$ ($\lambda \approx 0.25a$): (a) the non-dimensional acoustic radiation force in z-direction vs. $\phi_0$ (also means the location of object with respect to the wave source) for the case of solution for an elastic sphere active in $n=0$ and (b) the non-dimensional acoustic radiation force in z-direction vs. $\phi_0$ (also means the location of object with respect to the wave source) for the case of solution for a rigid sphere active in $n=0$. The dashed lines in both figures are showing the average of $Y_z$ over a complete wavelength.}
\label{Yzonemodeactivevsphi0ka4.1}
\end{figure}
Fig. (\ref{Yzonemodeactivevsphi0ka4.1}), is showing the non-dimensional acoustic radiation force in z-direction vs. $\phi_0/2\pi$ (which can be interpreted, the location of the object with respect to the wave source) for a complete wavelength at $ka=4.1$, in which, (a) is the result of implementation with piezo-elastic configuration and (b) is showing the result, when the formulation is derived in terms of surface velocity. The plots in Fig. (\ref{Yzonemodeactivevsphi0ka4.1}) confirms the explanations in section (\ref{AcousticRF}) that for scatterers active in a single mode, the average of acoustic radiation force over a complete wavelength is always positive and it is equal to the passive part of the force expression \cite{UltrasonicMojahed}. 
\begin{figure}[htp]
\centering
\includegraphics[scale=.22]{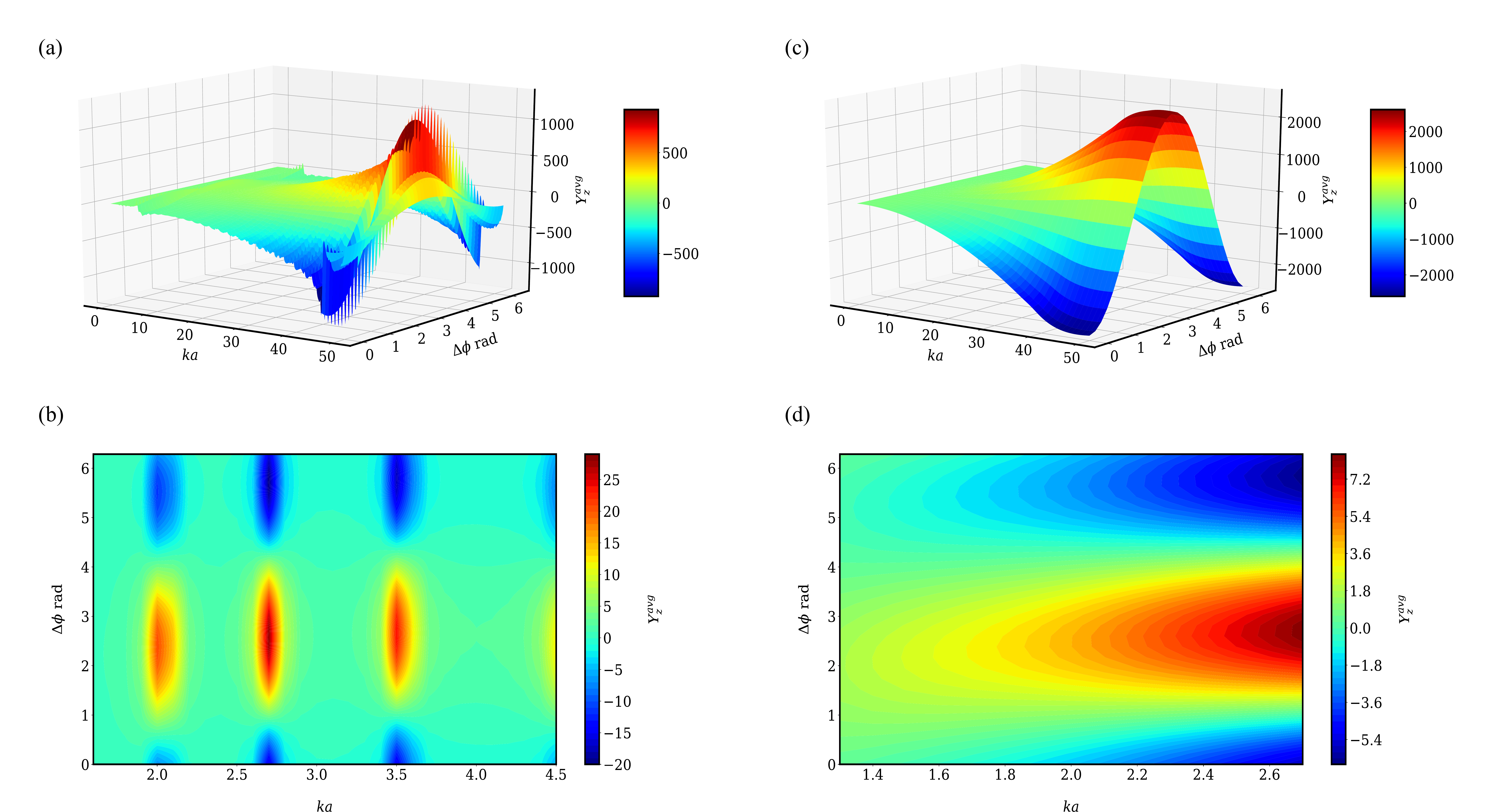}
\caption{The averaged acoustic radiation force in z-direction vs. the non-dimensional frequency and $\Delta\phi$ for (a) an elastic and (c) a rigid sphere activated in their zero an first modes.  }
\label{contours1}
\end{figure}
In order to show the practicality of the derived formulae, it is the time to discuss about figures which illustrate the average of force over the position of the object relative to the source (or its phase relative to the phase of incident wave) as a function of two efficacious parameters, frequency and the modal phase difference, defined as the phase difference between the vibrations of the shell in breathing and dipole modes. These parameters called efficacious as they are well-adjustable and if the results show wide intervals for the frequency and the phase-difference which the average of force is negative on those conditions, a practical system can be designed to work with a determined frequency and phase difference which brings about negative values for the average of force. Confidently, the desired occasion is happened for a wide range of practical frequencies and phase-differences which can be concluded from Fig. (\ref{contours1}). By setting $ka$ in the same operational range and considering $\Delta\phi$ in the range of $[0,2\pi]$ along with choosing the amplitudes of voltage or surface displacement from the results in Fig. (\ref{minvoltanddisp}), the averaged non-dimensional force (with respect to $\phi_0$) can be plotted vs. the non-dimensional frequency and the phase difference between the two modes of $n=0$ and $n=1$. Fig. (\ref{contours1}) is showing the 3D surfaces of averaged acoustic radiation force in z-direction vs. $ka$ and $\Delta\phi$ for (a) the solution which used a piezo-elastic solution and (c) the solution based on the surface velocity boundary condition. Both of the 3D surfaces are showing similar general trends in $ka$ and $\Delta\phi$ direction but the one for piezo-elastic solution has more fluctuations which a comparison between sub-plots (a) and (c) in Fig. (\ref{minvoltanddisp}), gives an idea of the reason for those fluctuations. Plots (b) and (d) are the contours of the 3D surfaces above them but in a more limited frequency range. The start non-dimensional frequency in plot (b) is chosen to be $ka \approx 2$ ($f\approx470$ kHz) as the minimum non-dimensional frequency at which a negative averaged force is observed is around $ka=2$. The start point for plot (d) is also chosen based on same criterion and it is, $ka\approx 1.2$ ($f\approx286$ kHz). The maximum non-dimensional frequencies in contours (b) and (d) are chosen arbitrarily.
\begin{figure}[htp]
\centering
\includegraphics[scale=.22]{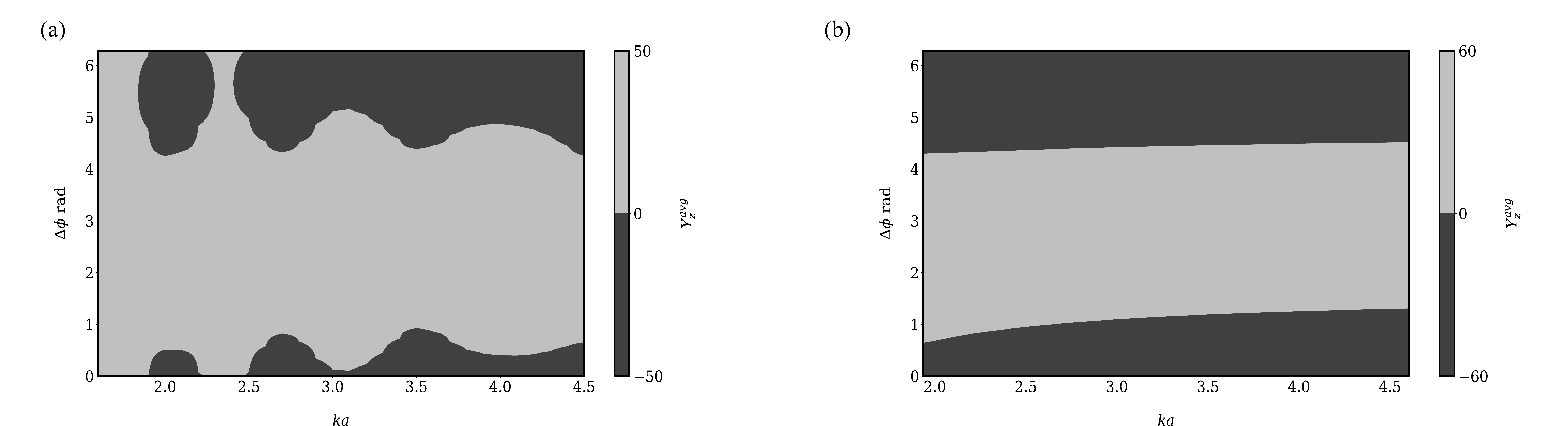}
\caption{(a) and (b) are showing the contours of averaged acoustic radiation force in z-direction vs. $ka$ and $\Delta\phi$ for the solution using voltage and surface displacement respectively. The darker islands are pointing the conditions under which the object experiences negative averaged force and the lighter islands are showing positive averaged force.  }
\label{contours2}
\end{figure}

The contours in sub-plots (b) and (d) of Fig. (\ref{contours1}) are plotted once more in Fig. (\ref{contours2}), but this time by using only two colors. The darker color in these contours is pointing the regions (pairs of $(ka,\Delta\phi)$) on which the averaged force is negative and the the lighter color is representing the conditions under which the averaged force is positive. The boundaries between two major colors are  showing the zero-averaged force states. Again, plots (a) and (b) are showing similarities in general trends but at the same time they illustrate the difference in the results when the equations are solved based on voltage of piezoelectric layer and when they are treated based on the surface velocity.

\begin{figure}[htp]
\centering
\includegraphics[scale=.22]{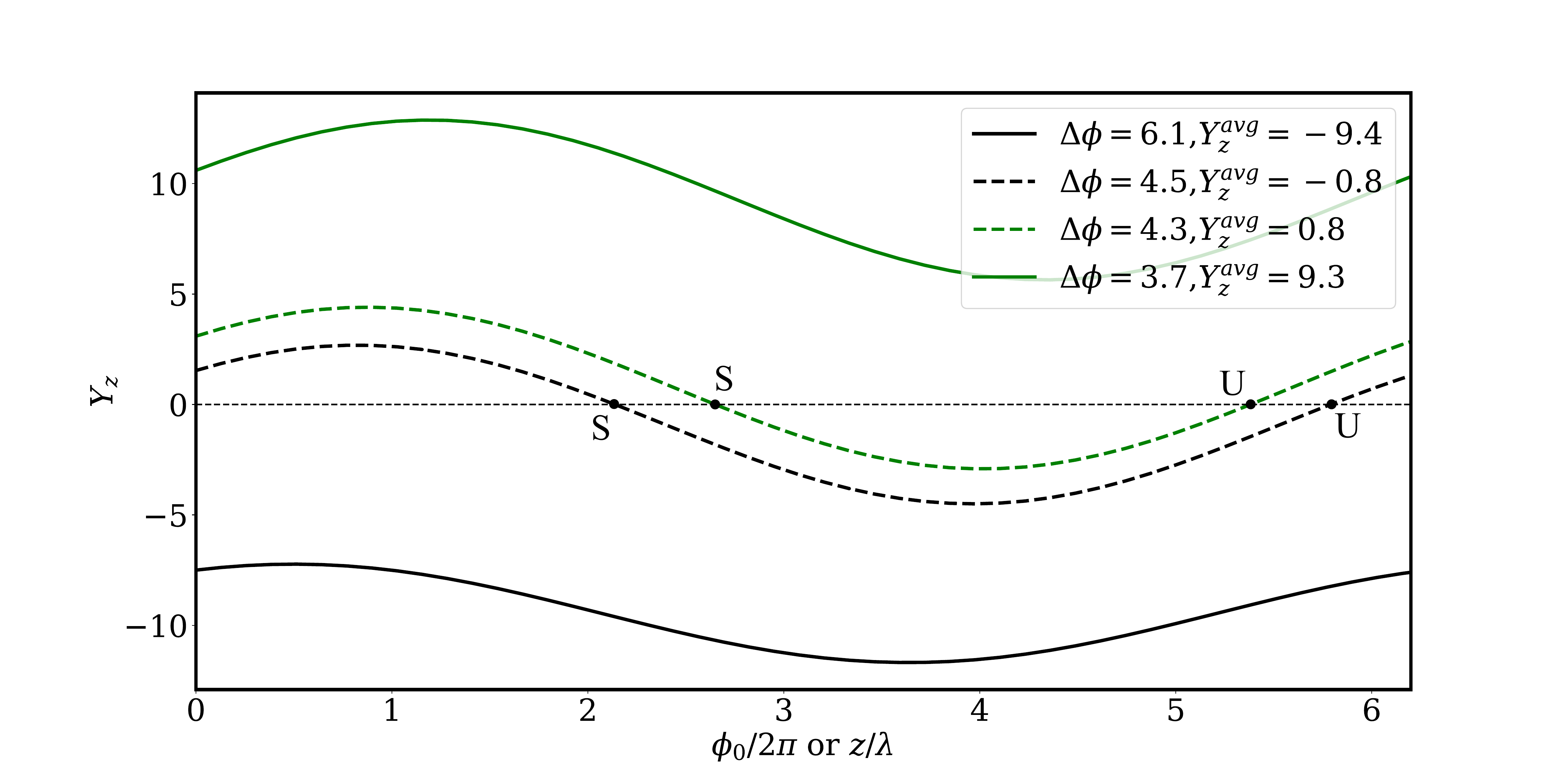}
\caption{Acoustic radiation force in z-direction at $ka\approx 3$ ($\lambda \approx1/3a$ ) for four different $\Delta\phi$. }
\label{4Forces}
\end{figure}
Keeping in mind the always positive averaged value of the acoustic radiation force over a
wavelength of the incident wave, $z/\lambda$, or over the phase difference between the incident wave
field and the radiated field of the carrier in its single mode of vibration (i.e., mono-pole or dipole
mode), $\phi_0/2\pi$ , this time for the case which the object is excited in two modes, Fig. (\ref{4Forces}) illustrates
the acoustic radiation force, $Y_z$, as a function of $z/\lambda$ or $\phi_0/2\pi$, for selected values the of modal
phase difference, $\Delta\phi$, at the selected frequency of $ka=3$. The sinusoidal patterns of the radiation
force function over the phase difference or spatial position  in Fig.(\ref{4Forces}), are generated in different states like big
positive averaged forces, big negative averaged forces and small positive or negative averaged
forces. According to Fig. (\ref{contours1}) it is understood that choice of the modal phase
difference at a specific frequency determines whether the averaged force is a positive number or
negative and also based on the magnitude of that averaged force one can guess whether the
radiation force only takes positive values (means always pushing state), only negative values
(means always pulling state) or like the previously reported case, it takes both positive and negative
values. The zero force state occurs in the last case and two rest states (i.e., $Y_z=0$) happens, one
of them as the stable rest state and the other as the unstable rest state, which both are indicated in Fig. (\ref{4Forces}) as S and U, respectively \cite{UltrasonicMojahed}.

\begin{figure}[htp]
\centering
\includegraphics[scale=.18]{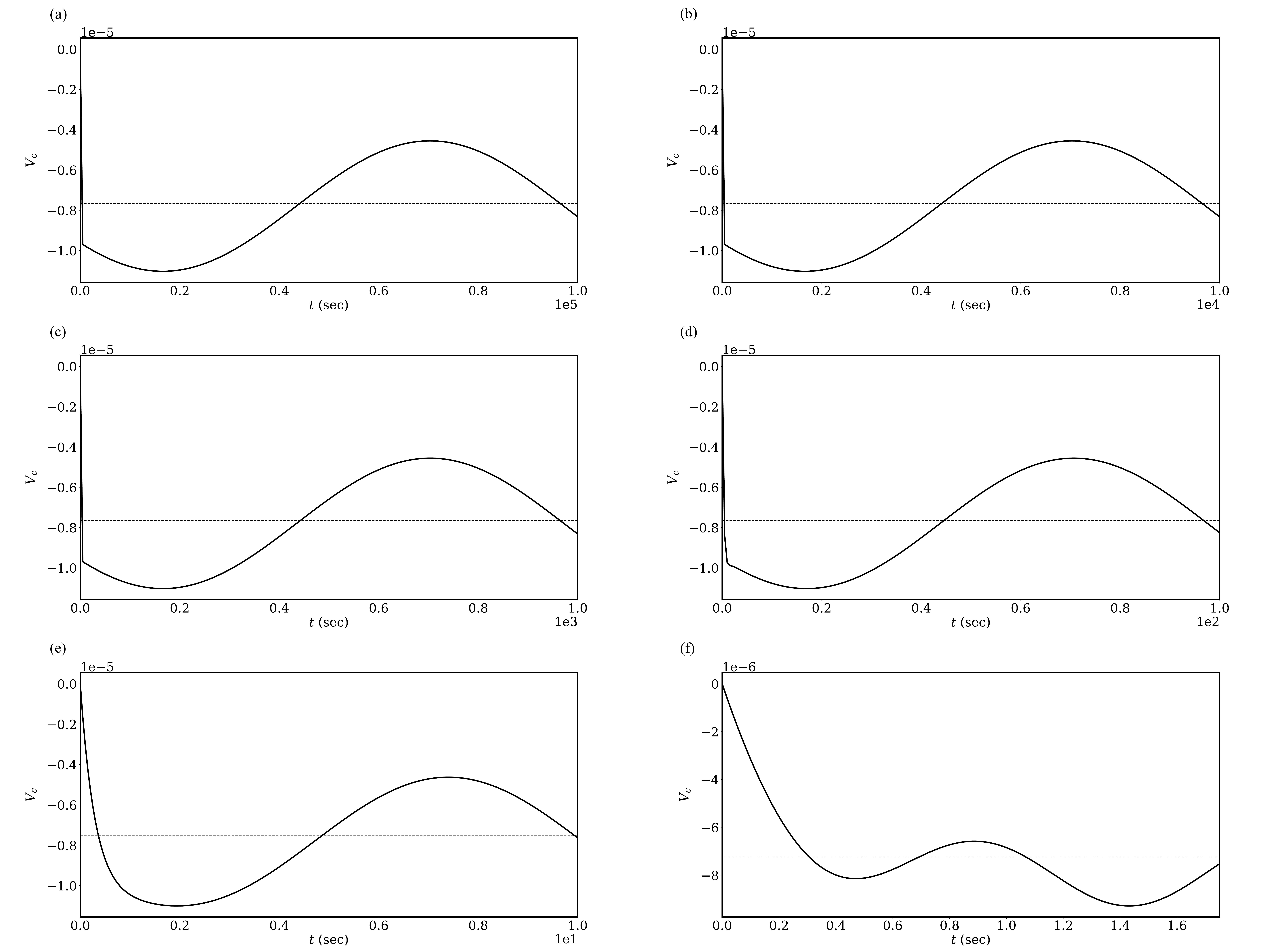}
\caption{The velocity of the carrier at $ka=2.7$ and $\Delta\phi=5.16$ rad vs. time for a sphere of outer radius $a=1$ mm with the velocity of phase change equal to (a) $0.01V_{passive}$, (b) $0.1V_{passive}$, (c) $V_{passive}$, (d) $10V_{passive}$, (e) $100V_{passive}$ and (f) $1000V_{passive}$. The dashed line is showing the velocity average of the carrier.}
\label{dynamics}
\end{figure}

After proving the possibility of generating the averaged negative acoustic radiation force state as well as the averaged positive states, with or without rest states, it is the time to put in challenge the suggested methodology of manipulation of activated carriers via the phase change (i.e., time varying) of the incident wave field. Obviously, the dynamics and swimming quality of the carrier should be considered. Fig. (\ref{dynamics}. a) through Fig. (\ref{dynamics}. f) illustrate the time variation of the velocity of the carrier, $V_c$, for different values of the rate of change of the incident wave phase, defined as $V_{phase}=d\phi_0/dt$ and selected as $V_{phase}/v_{passive}=0.01, 0.1, 1, 10, 100, 1000$ where $V_{passive}$ is the velocity of the carrier in its passive state calculated as $V_{phase}=S_cE_{inc}Y_{z}^{passive}/(6\pi\eta a)$. For simulation, it is assumed that the carrier is in the rest state, $V_c(t=0)=0$ m/s, the selected frequency is mapped to $ka=2.7$, and the selected modal phase difference is $\Delta\phi=5.61$ rad. As it is observed, for all the rates of incident wave phase change, the velocity of the carrier fluctuates around $S_cE_{inc}\overline{Y}_z/(6\pi\eta a)$ ($\overline{Y}_z$ is the force average in z-direction). The range of variation, $min(V_c)<V_c(t)<max(V_c)$,
is approximately the same for different values of $V_{phase}/V_{passive}$. The time scale of the carrier velocity
variation is related inversely to the phase change rate, as $\propto \left(V_{phase}/V_{passive}\right)^{-1}$. In conclusion, it is seen
that following the suggested strategy, the carrier is manipulated with fluctuating (i.e., sinusoidal
pattern) velocity around the pre-predicted velocity of swimming due to the induction of the
averaged radiation force in Stokes swimming condition, keeping in mind that the low Reynolds
number condition, $\rho_f a V_c/\eta \approx O(10^{-2})$, is valid.

\newpage
\section{Conclusions}\label{sec:conclusions}
The principal achievement of this paper consists of introducing a simple configuration to generate acoustic radiation force with a negative average, over one complete wavelength. In fact, a simple manipulation on the expression for the acoustic radiation force showed that a single mode active body can only experience negative force due to the interaction with a progressive acoustic wave but not a negative force average and the appearance of higher order trigonometric terms in the force expression are needed for a possible negative force average. Here, we put forward the idea of activating the body in zero and first modes simultaneously to generate negative force average and showed that considering a phase difference between the two modes of activation can act as a controlling parameter to achieve some desired conditions in the process of acoustic manipulation. In this paper, the machinery of the mentioned technique is done once by only considering the effect of activating body on the displacement of the surface of the spherical shell without addressing the process of activating the body, and once the problem been treated by using a piezo-elastic structure to simulate an example of the implementation of the method. It is notable that this work is trying to enrich the concepts and methods of complete manipulation of active carriers and practical concerns should be evaluated in future works.
\newpage
\section{Appendix}\label{sec:appendix}
\subsection{Coefficients in the Averaged Acoustic Radiation Force Formula}\label{appendix1}
In this appendix, we represent the full expressions for the coefficients in the formula for averaged acoustic radiation force in z-direction for the both cases of activation with surface velocity and with voltage. 
When the formulation is written based on the surface velocity of the sphere, the coefficients are:

\begin{align}
\label{Aeq}
\begin{split}\nonumber
     A=v_{amp}\bigg[2\cos\Delta\phi\bigg(\Re Z_2^1+&\Re Z_2^1\Re Z_1^0+\Im Z_1^0\Im Z_2^1+\alpha_2\Re Z_2^1+\beta_2\Im Z_2^1\bigg)+2\sin\Delta\phi\\\bigg(\Im Z_2^1+&\Im Z_2^1\Re Z_1^0-\Im Z_1^0\Re Z_2^1+\alpha_2\Im Z_2^1-\beta_2\Re Z_2^1\bigg)+\Re Z_2^0+2\Re Z_1^1\Re Z_2^0+2\Im Z_2^0 \Im Z_1^1\bigg]
\end{split}\\
\begin{split}
     B=v_{amp}\bigg[2\cos\Delta\phi\bigg(\Im Z_2^1+&\Im Z_2^1\Re Z_1^0-\Im Z_1^0\Re Z_2^1+\alpha_2\Im Z_2^1-\beta_2\Re Z_2^1\bigg)-2\sin\Delta\phi\\\bigg(\Re Z_2^1+&\Re Z_2^1\Re Z_1^0+\Im Z_1^0\Im Z_2^1+\alpha_2\Re Z_2^1+\beta_2\Im Z_2^1\bigg)+\Im Z_2^0+2\Re Z_1^1\Im Z_2^0-2\Re Z_2^0 \Im Z_1^1\bigg]
\end{split}
\end{align}

\begin{align}
\begin{split}\nonumber
     C=v_{amp}^2\bigg[2\cos\Delta\phi\bigg(\Im Z_2^1\Re Z_2^0+\Re Z_2^1\Im Z_2^0-\Im Z_2^0\Re Z_2^1-\Re Z_2^0\Im Z_2^1\bigg)+2\sin\Delta\phi\bigg(-\Re& Z_2^1\Re Z_2^0+\Im Z_2^1\Im Z_2^0\\-&\Im Z_2^0\Im Z_2^1+\Re Z_2^0\Re Z_2^1\bigg)\bigg]
\end{split}
\end{align}

\begin{align}
\begin{split}\nonumber
     D=E=v_{amp}^2\bigg[2\cos\Delta\phi\bigg(\Im Z_2^1\Im Z_2^0+\Re Z_2^1\Re Z_2^0\bigg)+2\sin\Delta\phi\bigg(-\Re Z_2^1\Im Z_2^0+\Im Z_2^1\Re Z_2^0\bigg)\bigg]
\end{split}
\end{align}

Also, the coefficients in the formula for the averaged acoustic radiation force, when the excitation is expressed using a piezo-elastic system is something similar to Eq. (\ref{Aeq}), but we do not mention it for to be brief. 
\subsection{Full Representation of Matrices for Piezo-elastic Structure}\label{appendix2}
In this appendix we represent the explicit expressions for some of the symbols used in part (\ref{piezoimplementation}). The vectors in Eq. (\ref{Constitutive}) are:

\begin{equation}
\begin{aligned}
    \label{vectorseq17}
    \boldsymbol{\sigma}=
    \begin{bmatrix}
\sigma_{rr} & \sigma_{\theta\theta} & \sigma_{\Theta\Theta}&\sigma_{r\theta}&\sigma_{r\Theta}&\sigma_{\theta\Theta}
\end{bmatrix}^T,\\
\boldsymbol{\Sigma}=
    \begin{bmatrix}
\Sigma_{rr} & \Sigma_{\theta\theta} & \Sigma_{\Theta\Theta}&\Sigma_{r\theta}&\Sigma_{r\Theta}&\Sigma_{\theta\Theta}
\end{bmatrix}^T,\\
\boldsymbol{s}=
    \begin{bmatrix}
s_{rr} & s_{\theta\theta} & s_{\Theta\Theta}&s_{r\theta}&s_{r\Theta}&s_{\theta\Theta}
\end{bmatrix}^T,\\
\boldsymbol{S}=
    \begin{bmatrix}
S_{rr} & S_{\theta\theta} & S_{\Theta\Theta}&S_{r\theta}&S_{r\Theta}&S_{\theta\Theta}
\end{bmatrix}^T,\\
\boldsymbol{\Delta}=
    \begin{bmatrix}
\Delta_{r} & \Delta_{\theta} & \Delta_{\Theta}
\end{bmatrix}^T\\
\end{aligned}
\end{equation}

and matrix operators in Eq. (\ref{Constitutive}) are like:

\begin{equation}
\begin{aligned}
    \label{sth}
    \boldsymbol{\Lambda}=
    \begin{bmatrix}
-\epsilon_{33}\nabla_2 & -\epsilon_{15}\frac{\partial}{\partial \theta} & -\frac{\epsilon_{15}}{\sin \theta}\frac{\partial}{\partial\Theta}
\end{bmatrix}^T,\\
\boldsymbol{\chi}=
\begin{bmatrix}
e_{33}\nabla_2 & e_{31}\nabla_2 & e_{31}\nabla_2 & e_{15}\frac{\partial}{\partial\theta}& \frac{e_{15}}{\sin\theta}\frac{\partial}{\partial\Theta}&0
\end{bmatrix}^T
\end{aligned}
\end{equation}
Also, the matrices of elasticity and piezo electricity can be shown like:

\begin{equation}
\begin{aligned}
    \label{vectorseq17}
\boldsymbol{c}=
    \begin{bmatrix}
c_{33} & c_{31} & c_{13}&0&0&0\\
c_{13} & c_{11} & c_{12}&0&0&0\\c_{13} & c_{12} & c_{11}&0&0&0\\0&0&0&2c_{44} & 0 & 0\\0&0&0&0 & 2c_{44} & 0\\0&0&0&0 & 0 & 2c_{66}
\end{bmatrix}\\
\boldsymbol{C}=
    \begin{bmatrix}
C_{33} & C_{31} & C_{13}&0&0&0\\
C_{13} & C_{11} & C_{12}&0&0&0\\C_{13} & C_{12} & C_{11}&0&0&0\\0&0&0&2C_{44} & 0 & 0\\0&0&0&0 & 2C_{44} & 0\\0&0&0&0 & 0 & 2C_{66}
\end{bmatrix}\\
\boldsymbol{e}=
    \begin{bmatrix}
e_{33} & e_{31} & e_{31}&0&0&0\\
0&0&0&2e_{15}&0&0\\0&0&0&0&2e_{15}&0
\end{bmatrix}
\end{aligned}
\end{equation}

In above representations, $e_{ij}$ are the constants in the piezoelectricity matrix, $\epsilon_{ij}$ are the dielectric constants,  $\nabla_2=r\frac{\partial}{\partial r}$ and $c_{ij}$, $C_{ij}$ are the elastic constants.\par
The $\mathbf{Q}$ in Eq. (\ref{EOM}) is:

\begin{equation}
    \label{Q}
    \mathbf{Q}=
\begin{bmatrix}
0 & \frac{\partial}{\partial\theta}+\cot\theta & -\cot\Theta&\nabla_2+2&0&\csc\theta\frac{\partial}{\partial\Theta}\\
0&0&\csc\theta\frac{\partial}{\partial\Theta}&2&\nabla_2&2\cot\Theta\\\nabla_2+1&-1&-1&\frac{\partial}{\partial\theta}+\cot\theta&\csc\theta\frac{\partial}{\partial\Theta}&0
\end{bmatrix}
\end{equation}
The modal functions $ {\mathcal{Z}}_{1}^{n}$ and $ {\mathcal{Z}}_{2}^{n}$ are as \cite{Mojahedrajabi1,rajabikhodavirdi1}:

\begin{equation}
\begin{aligned}
    \label{vectorseq17}
\mathcal{Z}_{1}^{n}=\frac{\begin{vmatrix}
S_{n}\left(1,3\right) & S_{n}\left(1,4\right) & S_{n}\left(1,5\right)&\frac{\omega}{c_{44}}\left(2n+1\right)\rho_f\varphi_0i^{n+1}j_n\left(ka\right)\\
S_{n}\left(2,3\right) & S_{n}\left(2,4\right) & S_{n}\left(2,5\right)&0\\S_{n}\left(4,3\right) & S_{n}\left(4,4\right) & S_{n}\left(4,5\right)&\frac{\varphi_0}{i\omega a}k\left(2n+1\right)\rho_f\varphi_0i^{n}j_n'\left(ka\right)\\T_{n}\left(6,3\right) & T_{n}\left(6,4\right) & T_{n}\left(6,5\right)&0
\end{vmatrix}}{\begin{vmatrix}
S_{n}\left(1,3\right) & S_{n}\left(1,4\right) & S_{n}\left(1,5\right)&-\frac{\omega}{c_{44}}\left(2n+1\right)\rho_f\varphi_0i^{n+1}h^{(1)}_n\left(ka\right)\\
S_{n}\left(2,3\right) & S_{n}\left(2,4\right) & S_{n}\left(2,5\right)&0\\S_{n}\left(4,3\right) & S_{n}\left(4,4\right) & S_{n}\left(4,5\right)&-\frac{\varphi_0}{i\omega a}k\left(2n+1\right)\rho_f\varphi_0i^{n}h^{(1)'}_n\left(ka\right)\\T_{n}\left(6,3\right) & T_{n}\left(6,4\right) & T_{n}\left(6,5\right)&0
\end{vmatrix}};\quad n>0\\
\mathcal{Z}_{2}^{n}=\frac{\begin{vmatrix}
S_{n}\left(1,3\right) & S_{n}\left(1,4\right) & S_{n}\left(1,5\right)&0\\
S_{n}\left(2,3\right) & S_{n}\left(2,4\right) & S_{n}\left(2,5\right)&0\\S_{n}\left(4,3\right) & S_{n}\left(4,4\right) & S_{n}\left(4,5\right)&0\\T_{n}\left(6,3\right) & T_{n}\left(6,4\right) & T_{n}\left(6,5\right)&1
\end{vmatrix}}{\begin{vmatrix}
S_{n}\left(1,3\right) & S_{n}\left(1,4\right) & S_{n}\left(1,5\right)&-\frac{\omega}{c_{44}}\left(2n+1\right)\rho_f\varphi_0i^{n+1}h^{(1)}_n\left(ka\right)\\
S_{n}\left(2,3\right) & S_{n}\left(2,4\right) & S_{n}\left(2,5\right)&0\\S_{n}\left(4,3\right) & S_{n}\left(4,4\right) & S_{n}\left(4,5\right)&-\frac{\varphi_0}{i\omega a}k\left(2n+1\right)\rho_f\varphi_0i^{n}h^{(1)'}_n\left(ka\right)\\T_{n}\left(6,3\right) & T_{n}\left(6,4\right) & T_{n}\left(6,5\right)&0
\end{vmatrix}};\quad n>0
\end{aligned}
\end{equation}
where, $S_n(i,j)$ are the elements of the global structural transfer matrix \cite{Mojahedrajabi1} which its derivation for a single layer is presented in the next appendix. Also, the expressions for $\mathcal{Z}_{1}^{0}$ and $\mathcal{Z}_{2}^{0}$ are like those represented for $n>0$ but the second row of each determinant should be eliminated \cite{Mojahedrajabi1}.

\subsection{Formulation on the general theory of elasticity for a sphere}\label{elastodyn}
In this appendix we review a method of finding scattering coefficients of an elastic spherical shell using the full equations of elasticity. For an spherically isotropic medium, the governing equations of motion, constitutive equations and the strain-displacement relations can be represented like \cite{scandrett2002scattering}:

\begin{equation}
    \label{EOMsphere}
    \begin{aligned} \boldsymbol{\Gamma}\boldsymbol{\sigma}=\rho\ddot{\boldsymbol{u}}\\ \boldsymbol{\sigma}=\boldsymbol{CS}\\
    \boldsymbol{S}=\boldsymbol{H}\boldsymbol{u}
    \end{aligned}
\end{equation}
where, $\boldsymbol{u}=\left[u_r,u_\theta,u_\phi\right]$ is the displacement vector, $\boldsymbol{\sigma}=\left[\sigma_{rr},\sigma_{\theta\theta},\sigma_{\phi\phi},\sigma_{r\theta},\sigma_{r\phi},\sigma_{\theta\phi}\right]$ is the stress vector, $\boldsymbol{S}=\left[S_{rr},S_{\theta\theta},S_{\phi\phi},S_{r\theta},S_{r\phi},S_{\theta\phi}\right]$ is the strain vector and $\boldsymbol{C}$ is the stiffness matrix for isotropic materials with 12 non-zero elements and only 2 independent variables. The matrices $\boldsymbol{\Gamma}$ and $\boldsymbol{H}$ are so well-known and we do not repeat them here with the aim of brevity \cite{hu1954general,ding1996natural}. In the current theory, the displacement is divided into two classes. In the first class, both of radial displacement and dilatation are zero and for the second one, the radial curl of the displacements is equal to zero. To reduce the order and the coupling of current PDEs a method of change of variables is introduced which substitutes the displacements and two of shear stresses with new functions called displacement and stress potentials of $w,g,\psi,\sigma_1$ and $\sigma_2$  \cite{hu1954general}:

\begin{equation}
    \label{ChangeofVars}
    \begin{aligned} 
    u_r=w,\quad u_\theta=-\frac{1}{\sin\theta}\frac{\partial\psi}{\partial\phi}-\frac{\partial g}{\partial\theta}, \quad u_\phi=-\frac{1}{\sin\theta}\frac{\partial g}{\partial\phi}+\frac{\partial\psi }{\partial\theta}\\   \sigma_{r\theta}=-\frac{1}{\sin\theta}\frac{\partial\sigma_1}{\partial\phi}-\frac{\partial \sigma_2}{\partial\theta}, \quad \sigma_{r\phi}=-\frac{1}{\sin\theta}\frac{\partial \sigma_2}{\partial\phi}+\frac{\partial\sigma_1 }{\partial\theta}
    \end{aligned}
\end{equation}
Using the introduced change of variables, all the other terms like $\sigma_{\theta\theta}$, $\sigma_{\theta\phi}$ and $\sigma_{\phi\phi}$ can be re-written in terms of new variables. One can also use the symmetries in the current geometry and expand all the variables in terms of spherical harmonics like:
\begin{equation}
    \label{harmonics}
    \psi(r)=\sum_{n}\psi^{n}P_n(\cos\theta)
\end{equation}
to eliminate all the angular partial derivatives. After some lengthy manipulations, two sets of normalized first order ordinary differential equations are obtained like:

\begin{equation}
    \label{StateSpace}
    \begin{aligned} 
    \frac{d}{d\xi}\boldsymbol{v}_1^n(\xi,\omega)=\boldsymbol{m}_1^n\boldsymbol{v}_1^n(\xi,\omega)\\  
    \frac{d}{d\xi}\boldsymbol{v}_2^n(\xi,\omega)=\boldsymbol{m}_2^n\boldsymbol{v}_2^n(\xi,\omega)
    \end{aligned}
\end{equation}

where, $n=(1,2,...)$,  $\boldsymbol{v}_1^n=\left[\sigma_1^n,\psi^n\right]$, $\boldsymbol{v}_2^n=\left[\sigma_{rr}^n,\sigma_{2}^n,g^n,w^n\right]^T$, $\xi=r/a$ and $a$ is the outer radius of the shell. For $n=0$, as the field variables in Eq. (\ref{ChangeofVars}) are definable only by the derivatives of $\psi, g, \sigma_1$ and $\sigma_2$, their zero mode coefficients can be considered as zero without loss of generality and the two equations in Eq. (\ref{StateSpace}) collapse to:
\begin{equation}
    \label{StateSpace}
    \frac{d}{d\xi}\boldsymbol{v}_2^0(\xi,\omega)=\boldsymbol{m}_1^0\boldsymbol{v}_2^0(\xi,\omega)\\ 
\end{equation}

where, $\boldsymbol{v}_2^n=\left[\sigma_{rr}^0,w^0\right]^T$. For all the different scenarios like a free vibrating shell or a radiating shell due to the impact of an incident wave, the ODEs in Eq. (\ref{StateSpace}) can be solved through various of methods. Here, we employ the method of Neumann series to find the fundamental matrix of the current system of ODEs \cite{boyce2021elementary,chen2001free}. According to Taylor expansion, and by assuming the continuity of all the variables through the elastic shell, one can write the following:
\begin{equation}
    \label{Taylor}
    \boldsymbol{v}_i^n(\xi)=\boldsymbol{v}_i^n(0)+\frac{\boldsymbol{v}_i^{n'}(0)}{1!}\xi+\frac{\boldsymbol{v}_i^{n''}(0)}{2!}\xi^2+...
\end{equation}
where, $i=1,2$ and this expression is showing the value of $\boldsymbol{v}_i^n$ at a neighbouring point of $r=0$,like $r=\xi$, and all the derivatives can be generated from Eq. (\ref{StateSpace}). For a better convergence of this method, we divide the whole shell to adjacent laminates, and as a result one can write:
\begin{equation}
    \label{TransferMatrix}
    \boldsymbol{v}_i^n(\xi=\frac{pd}{a}-\frac{b}{a})=\boldsymbol{S}_i^n\boldsymbol{v}_i^n(\xi=\frac{b}{a})
\end{equation}
where, $b$ is the inner radius of the shell, $p$ is the number of laminates, $d$ is the width of the laminates,   $\boldsymbol{S}_i^n=\Pi_{j=1}^p\boldsymbol{A}_{ij}^n(\xi_j)$ and $\boldsymbol{A}_{ij}^n(\xi)=exp(\xi\boldsymbol{m}_{i}^n)$ \cite{chen2001free}. For a single elastic layer, the following boundary conditions can be considered: $\sigma_2(r=a,b)=\sigma_1(r=a,b)=0, \sigma_{rr}(r=a)=-p_{f}(r=a), \sigma_{rr}(r=b)=0, w(r=a)=\frac{v_f}{-i\omega}$, where $p_f$ and $v_f$ are total pressure field and velocity of the fluid respectively. For the case of piezo-elastic structure in this paper, the appropriate boundary conditions should be applied at the interface of the elastic layer and the environment, between the two layers and at the internal layer of the piezo. They are: I. no tangential stress component at the inner/outer surface of the piezo/elastic layer ($\sigma_2(r=a)=\Sigma_2(r=c)=0$), II. continuity of the normal stress component at the inner and outer surface, ($\sigma_{rr}(r=a)=-P$, $\Sigma_{rr}(r=c)=0$), where $P$ is the total normal pressure of the fluid, III. continuity of traction and displacement vectors at the interface of piezo and elastic layer ($r=b$), and IV. Continuity of the normal component of the velocity in fluid and the one in solid at the interface of elastic and fluid, ($r=a$). 

\subsection{Material Properties}\label{MP}
The material properties for the elastic and the PZT4 layer are mentioned in this appendix. For the PZT4 layer, $\rho_p=7500$ Kg/$m^3$, $C_{11}=139$ GPa, $C_{12}=78$ GPa, $C_{13}=74.3$ GPa, $C_{33}=115$ GPa, $C_{44}=25.6$ GPa, $C_{66}=30.5$ GPa, $\epsilon_{11}=650$ $10^{-11}$F/m, $\epsilon_{33}=560$ $10^{-11}$F/m. For the elastic material we used, $\rho_p=7850$ Kg/$m^3$, $E=207$ GPa, and $\nu=0.29$.

\section*{References}

%

\end{document}